\documentclass{article}
\usepackage{amssymb}
\usepackage{amsmath}

\input{tcilatex}

\begin{document}

\title{Why the Decision-Theoretic Perspective Misrepresents Frequentist
Inference: `Nuts and Bolts' vs. Learning from Data}
\author{Aris Spanos \\
Department of Economics,\\
Virginia Tech, USA}
\date{August 2016}
\maketitle

\begin{abstract}
The primary objective of this paper is to revisit and make a case for the
merits of R.A. Fisher's objections to the decision-theoretic framing of
frequentist inference. It is argued that this framing is congruent with the
Bayesian but incongruent with the frequentist inference. It provides the
Bayesian approach with a theory of optimal inference, but it misrepresents
the theory of optimal frequentist inference by framing inferences solely in
terms of the universal quantifier `for all values of $\theta $ in $\Theta $%
', denoted by $\forall \theta \mathit{\in }\Theta $. This framing is at odds
with the primary objective of model-based frequentist inference, which is to
learn from data about the true value $\theta ^{\ast };$ the one that gave
rise to the particular data. The frequentist approach relies on factual
(estimation, prediction), as well as hypothetical (testing) reasoning, both
of which revolve around the existential quantifier $\exists \theta ^{\ast }%
\mathit{\in }\Theta $. The paper calls into question the appropriateness of
admissibility and reassesses Stein's paradox as it relates to the capacity
of frequentist estimators to pinpoint $\theta ^{\ast }$. The paper also
compares and contrasts loss-based errors with traditional frequentist
errors, such as coverage, type I and II; the former are attached to $\theta ,
$ but the latter to the inference procedure itself.\medskip 

\end{abstract}

\newpage 

\section{Introduction\protect\vspace*{-0.06in}}

Wald's (1950) decision-theoretic framework is widely viewed as providing a
broad enough perspective to accommodate and compare the frequentist and
Bayesian approaches to inference, despite their well-known differences. It
is perceived as offering a neutral framing of inference that brings into
focus their common features and tones down their differences; see Berger
(1985), Robert (2001), O'Hagan (1994).

Historically, Wald (1939) proposed the original variant of the
decision-theoretic framework with a view to unify Neyman's (1937) rendering
of frequentist interval estimation and testing:

\textsf{\textquotedblleft The problem in this formulation is very general.
It contains the problems of testing hypotheses and of statistical estimation
treated in the literature.\textquotedblright\ (p. 340)}

Among the frequentist pioneers, Jerzy Neyman accepted enthusiastically this
broader perspective, primarily because the concepts of decision rules and
action spaces seemed to provide a better framing for his \textit{%
behavioristic interpretation} of Neyman-Pearson (N-P) testing based on the
accept/reject rules; see Neyman (1952; 1971). Neyman's attitude towards
Wald's (1950) framing was also adopted wholeheartedly by some of his most
influential students/colleagues at Berkeley, including Lehmann (1959) and
LeCam (1986). In a foreword of a collection of Neyman's early papers, his
students/editors described the Wald's framing as \textsf{\ }(Neyman, 1967,
p. vii):

\textsf{\textquotedblleft A natural but far reaching extension of their [N-P
formulation] scope can be found in Abraham Wald's theory of statistical
decision functions.\textquotedblright}

At the other end of the argument, R. A. Fisher (1955) rejected Wald's
framing on the grounds that it seriously distorts his rendering of
frequentist statistics:

\hspace*{-0.25in}\textsf{\textquotedblleft The attempt to reinterpret the
common tests of significance used in scientific research as though they
constituted some kind of acceptance procedure and led to `decisions' in
Wald's sense, originated in several misapprehensions and has led,
apparently, to several more.\textquotedblright\ (p. 69)}

With a few exceptions, such as Cox (1958), Tukey (1960) and Birnbaum (1977),
Fisher's (1955) viewpoint has been inadequately discussed and evaluated by
the subsequent statistics literature. The primary aim of this paper is to
revisit Fisher's minority view by taking a closer look at the
decision-theoretic framework with a view to re-evaluate the claim that it
provides a neutral framework for comparing the frequentist and Bayesian
approaches. It is argued that Fisher's view that the decision theoretic
framing is germane to "acceptance sampling", but misrepresents frequentist
inference, is not without merit. The key argument of the discussion that
follows is that the decision-theoretic notions of loss function and
admissibility are congruent with the Bayesian approach, but incongruent with
both the primary objective and the underlying reasoning of the frequentist
approach.

Section 2 introduces the basic elements of the decision theoretic set-up
with a view to bring out its links to the Bayesian and frequentist
approaches, calling into question the conventional wisdom concerning its
neutrality. Section 3 takes a closer look at the Bayesian approach and
argues that had the decision-theoretic apparatus not exist, Bayesians would
have been forced to invent it in order to establish a theory of optimal
Bayesian inference. Section 4 discusses critically the notions of loss
functions and admissibility, focusing primarily on their role in giving rise
to Stein's paradox and their incompatibility with the frequentist approach.
It is argued that the frequentist dimension of the notions of a loss
function and admissibility is more apparent than real. Section 5 makes a
case that the decision-theoretic framework misrepresents both the primary
objective and the underlying reasoning of the frequentist approach. Section
6 revisits the notion of a loss function and its dependence on `information
other than the data'. It is argued that loss-based errors are both different
and incompatible with the traditional frequentist errors because they are
attached to the unknown parameters instead of the inference procedures
themselves.\vspace*{-0.15in}

\section{\textbf{The decision theoretic set-up}\protect\vspace*{-0.12in}}

\subsection{\textbf{Basic elements of the decision-theoretic framing}\protect%
\vspace*{-0.06in}}

The current decision-theoretic set-up has three basic elements.

\textbf{1}. A prespecified (parametric) statistical model $\mathcal{M}_{%
\mathbf{\theta}}(\mathbf{x})$, generically specified by:$\vspace*{-0.06in}$ 
\begin{equation}
\begin{array}{c}
\mathcal{M}_{\mathbf{\theta}}(\mathbf{x})\mathbf{=}\{f(\mathbf{x};\mathbf{%
\theta}),\ \mathbf{\theta}\mathit{\in}\Theta\},\ \mathbf{x}\mathit{\in}%
\mathbb{R}_{X}^{n},\text{ for }\mathbf{\theta}\mathit{\in}\Theta\mathbf{%
\subset}\mathbb{R}^{m},\ m\mathit{\ll}n,%
\end{array}
\vspace*{-0.06in}  \label{smo}
\end{equation}
where $f(\mathbf{x};\mathbf{\theta})$ denotes the (joint)\textit{\
distribution of the sample} $\mathbf{X}$\textbf{:}$\mathbf{=}%
(X_{1},...,X_{n})\mathbf{,}$ $\mathbb{R}_{X}^{n}$ denotes the \textit{sample
space} and $\Theta$ the \textit{parameter space}. This model represents the
stochastic mechanism assumed to have given rise to data $\mathbf{x}_{0}$%
\textbf{:}$\mathbf{=}(x_{1},...,x_{n}).\smallskip$

\textbf{2}. A decision space $D$ containing all mappings $d(.)$: $\mathbb{R}%
_{X}^{n}\rightarrow A,$ where $A$ denotes the set of all actions available
to the statistician.$\smallskip$

\textbf{3}. A loss function $L(.,.)\mathit{:}$ $[D\times\Theta]\rightarrow
R, $\ representing the numerical loss if the statistician takes action $a%
\mathit{\in}A$\ when the state of nature is $\theta\mathit{\in}\Theta;$ see
Ferguson (1967), Berger (1985), Wasserman (2004).$\smallskip$

The basic idea is that, when the decision-maker selects action $a$, he/she
does not know the `true' state of nature, represented by $\mathbf{\theta }%
^{\ast}.$ However, contingent on each action $a\mathit{\in}A,$ the
decision-maker `knows' the losses (gains, utilities) resulting from
different choices $\left( d,\mathbf{\theta}\right) \mathit{\in}%
[D\times\Theta].$ The decision-maker observes data $\mathbf{x}_{0},$ which
provides some information about $\mathbf{\theta}^{\ast}\mathbf{,}$ and then
maps each $\mathbf{x}\mathit{\in}\mathbb{R}_{X}^{n}$ to a certain action $a%
\mathit{\in}A$ guided solely by $L(d,\mathbf{\theta}).$\vspace*{-0.15in}

\subsection{The original Wald framing\protect\vspace*{-0.06in}}

It is important to bring out the fact that the original Wald (1939) framing
was much narrower than the above (1-3) rendering, due to its aim to
formalize the Neyman-Pearson (N-P) approach; see Ferguson (1976). What were
the key differences?

(i) The decision (action) space $D$ was defined exclusively in terms of
subsets of the parameter space $\Theta$. For estimation $D\mathit{:=}%
\{\theta:\theta\mathit{\in}\Theta\}$ is the set of all singleton points of $%
\Theta$ and for testing $D\mathit{:=}\left( \Theta_{0},\Theta_{1}\right) $,
the null and alternative regions, respectively.

(ii) The original loss (weight) was a zero-positive function, with zero loss
at $\theta \mathit{=}\theta ^{\ast }$:$\vspace*{-0.06in}$%
\begin{equation}
L_{0-c}(\theta ,\widehat{\theta }(\mathbf{X}))\mathbf{=}\left\{ 
\begin{array}{l}
0\text{ if }\widehat{\theta }(\mathbf{X})=\theta ^{\ast } \\ 
c_{\theta }\mathit{>}0\text{ if }\widehat{\theta }(\mathbf{X})\mathit{=}%
\theta \mathit{\neq }\theta ^{\ast },\ \theta \mathit{\in }\Theta ,%
\end{array}%
\right. \vspace*{-0.06in}  \label{w}
\end{equation}%
where $\theta ^{\ast }$ is the true value of $\theta $ in $\Theta .$ For the
discussion that follows it is important to note that (\ref{w}) is
non-operational in practice because $\theta ^{\ast }$ is unknown.

The more general framing, introduced by Wald (1947; 1950) and broadened by
Le Cam (1955), extended the scope of the original set-up by generalizing the
notions of loss functions and decision spaces. In what follows it is argued
that these extensions created serious incompatibilities with both the
objective and the underlying reasoning of frequentist inference.

In addition, it is both of historical and methodological interest to note
that Wald (1939) introduced the notion of a prior distribution $\pi(\theta),$
$\forall\theta\mathit{\in}\Theta,$ into the original decision-theoretic
machinery reluctantly, and justified it on being a useful tool for proving
certain theorems:

\textsf{\textquotedblleft The situation regarding the introduction of an a
priori probability distribution of }$\theta$\textsf{\ is entirely different.
First, the objection can be made against it, as Neyman has pointed out, that 
}$\theta$\textsf{\ is merely an unknown constant and not a variate, hence it
makes no sense to speak of the probability distribution of }$\theta$\textsf{%
. Second, even if we may assume that }$\theta$\textsf{\ is a variate, we
have in general no possibility of determining the distribution of }$\theta $%
\textsf{\ and any assumptions regarding this distribution are of
hypothetical character. The reason why we introduce here a hypothetical
probability distribution of }$\theta$\textsf{\ is simply that it proves to
be useful in deducing certain theorems and in the calculation of the best
system of regions of acceptance.\textquotedblright\ (p. 302)}

He was also\ the\ first\ to\ highlight\ the\ extreme\ relativism of the
decision-theoretic\ notion of `optimality' with respect a particular loss
function:

\textsf{\textquotedblleft The "best" system of regions of acceptance ...
will depend \textit{only} on the weight function of the
errors.\textquotedblright\ (p. 302) [emphasis added]}\vspace*{-0.15in}

\subsection{\textbf{A shared neutral framework?}\protect\vspace*{-0.06in}}

The frequentist, Bayesian and the decision-theoretic approaches share the
notion of a statistical model by viewing data $\mathbf{x}_{0}\mathit{:=}%
(x_{1},...,x_{n})$ as a realization of a sample $\mathbf{X}\mathit{:=}%
(X_{1},...,X_{n})$ from (\ref{smo}).

The key differences between the three approaches are:

(a) the frequentist approach relies exclusively on $\mathcal{M}_{\mathbf{%
\theta}}(\mathbf{x}),$

(b) the Bayesian approach adds a \textit{prior} distribution,\ $\pi (\mathbf{%
\theta}),\ \forall\mathbf{\theta}\mathit{\in}\Theta\ $(for all $\theta%
\mathit{\in}\Theta$),

(c) the decision-theoretic framing revolves around a \textit{loss} (gain or
utility) function:\vspace*{-0.06in}%
\[
\begin{array}{c}
L(d(\mathbf{x}),\mathbf{\theta}),\forall\mathbf{\theta}\mathit{\in}\Theta,\
\forall\mathbf{x}\mathit{\in}\mathbb{R}_{X}^{n}.%
\end{array}
\vspace*{-0.06in}
\]
The loss function is often assumed to be an even, differentiable and convex
function of $(d(\mathbf{x})\mathit{-}\mathbf{\theta}),$ and can take
numerous functional forms; see Wasserman (2004), Robert (2001), Bansal
(2007) inter alia.

The claim that the decision-theoretic perspective provides a neutral ground
is often justified on account of the loss function\textbf{\ }being a
function of the sample and parameter spaces via the two \textit{universal
quantifiers:}

(i) `$\forall\mathbf{x}\mathit{\in}\mathbb{R}_{X}^{n}$', associated with the 
\textit{distribution of the sample}:\vspace*{-0.06in}%
\[
\begin{tabular}{ll}
\textbf{frequentist}: & $f(\mathbf{x};\mathbf{\theta}),\ \forall \mathbf{x}%
\mathit{\in}\mathbb{R}_{X}^{n},$%
\end{tabular}
\vspace*{-0.06in}
\]

(ii) `$\forall \mathbf{\theta }\mathit{\in }\Theta $' associated with the 
\textit{posterior distribution}:\vspace*{-0.06in}%
\[
\begin{tabular}{ll}
\textbf{Bayesian}: & $\pi (\mathbf{\theta \mathit{\mid }x}_{0})\mathit{=}%
\frac{\pi (\mathbf{\theta })\cdot f(\mathbf{x}_{0}\mathit{\mid }\mathbf{%
\theta })}{\int_{\mathbf{\theta }\mathit{\in }\Theta }\pi (\mathbf{\theta }%
)\cdot f(\mathbf{x}_{0}\mathit{\mid }\mathbf{\theta })d\mathbf{\theta }},\
\forall \mathbf{\theta }\mathit{\in }\Theta .$%
\end{tabular}%
\ \ \ \vspace*{-0.06in}
\]%
The idea is that allowing for all values of $\mathbf{x}\mathit{\ }$in $%
\mathbb{R}_{X}^{n}$ goes beyond the Bayesian perspective, which relies
exclusively on a single point $\mathbf{x}_{0}.~$What is not obvious is
whether that is sufficient to do justice to the frequentist approach. A
closer scrutiny suggests that frequentist inference is misrepresented by the
way both quantifiers are used in the decision-theoretic framing of inference.

First, the quantifier $\forall \mathbf{x}\mathit{\in }\mathbb{R}_{X}^{n}$
plays a minor role since its only relevance is in defining the expectation
for transforming a loss function, say $L(\theta ,\widehat{\theta }(\mathbf{x}%
)),$ into a risk function:$\vspace*{-0.06in}$%
\begin{equation}
\begin{array}{c}
R(\theta ,\widehat{\theta })\mathbf{=}E_{\mathbf{X}}\left[ L(\theta ,%
\widehat{\theta }(\mathbf{X}))\right] \mathbf{=}\int_{\mathbf{x}\in \mathbb{R%
}_{X}^{n}}L(\theta ,\widehat{\theta }(\mathbf{x}))f(\mathbf{x};\theta )d%
\mathbf{x,}\text{ }\forall \theta \mathit{\in }\Theta .%
\end{array}%
\vspace*{-0.06in}  \label{rf}
\end{equation}%
This is the only place where the underlying statistical model $\mathcal{M}_{%
\mathbf{\theta }}(\mathbf{x})$ enters the decision-theoretic framing. Hence,
from this perspective the only relevant part of the behavior of $\widehat{%
\theta }(\mathbf{X})$ is how it affects the risk for different values of $%
\theta $ in $\Theta .$ This undermines the pivotal role of the quantifier $%
\forall \mathbf{x}\mathit{\in }\mathbb{R}_{X}^{n}$ in determining the theory
of optimal frequentist inference. For that the distribution of the sample, $%
f(\mathbf{x};\theta ),\ \forall \mathbf{x}\mathit{\in }\mathbb{R}_{X}^{n},$
takes center stage since the sampling distribution of any statistic $Y_{n}%
\mathbf{=}g\mathbf{(X)}$\textbf{\ }(estimator, test, predictor) is derived
via:\vspace*{-0.1in}%
\begin{equation}
\begin{array}{c}
F(y;\theta )\mathit{:=}\mathbb{P}(Y_{n}\mathit{\leq }y;\theta )\mathit{=}%
\underbrace{\int \int \cdot \cdot \cdot \int }_{\{\mathbf{x}:\ g(\mathbf{x}%
)\leq t;\ \mathbf{x}\mathit{\in }\mathbb{R}_{X}^{n}\}}f(\mathbf{x};\theta )d%
\mathbf{x}.%
\end{array}%
\vspace*{-0.1in}  \label{f}
\end{equation}%
The relevant error probabilities that calibrate the optimality of
frequentist procedures are defined in terms of (\ref{f}).

Second, the decision-theoretic notion of optimality revolves around the
universal quantifier `$\forall \mathbf{\theta }\mathit{\in }\Theta $',
rendering it congruent with the Bayesian but incongruent with the
frequentist approach. To be more specific, since different risk functions
often intersect over $\Theta ,$ an optimal rule is usually selected after
the risk function is reduced to a scalar. Two such choices of relevant risk
are:$\vspace*{-0.07in}$%
\begin{equation}
\begin{tabular}{rl}
\textbf{Maximum risk}: & $R_{\max }(\widehat{\mathbf{\theta }})\mathbf{=}%
\underset{\mathbf{\theta }\mathit{\in }\Theta }{\sup }R(\mathbf{\theta },%
\widehat{\mathbf{\theta }}),\medskip $ \\ 
\textbf{Bayes risk}: & $R_{B}(\widehat{\mathbf{\theta }})\mathbf{=}\int_{%
\mathbf{\theta }\in \Theta }R(\mathbf{\theta },\widehat{\mathbf{\theta }}%
)\pi (\mathbf{\theta })d\mathbf{\theta }.$%
\end{tabular}%
\ \vspace*{-0.06in}  \label{ri}
\end{equation}%
Hence, an obvious way to choose among different rules is to find the one
that minimizes the relevant risk with respect to all possible estimates $%
\widetilde{\theta }(\mathbf{x}).$ In the case of (\ref{ri}), this gives rise
to two corresponding decision rules:$\vspace*{-0.07in}$%
\[
\begin{tabular}{rl}
\textbf{Minimax rule}: & $\underset{\widetilde{\theta }(\mathbf{x})}{\inf }%
R_{\max }(\widehat{\theta })\mathbf{=}\underset{\widetilde{\theta }(\mathbf{x%
})}{\inf }[\underset{\theta \mathit{\in }\Theta }{\sup }R(\theta ,\widehat{%
\theta })],\medskip $ \\ 
\textbf{Bayes rule}: & $\underset{\widetilde{\theta }(\mathbf{x})}{\inf }%
R_{B}(\widehat{\theta })\mathbf{=}\underset{\widetilde{\theta }(\mathbf{x})}{%
\inf }\int_{\theta \in \Theta }R(\theta ,\widehat{\theta })\pi (\theta
)d\theta .$%
\end{tabular}%
\ \vspace*{-0.1in}
\]%
In this sense, a decision or a Bayes rule $\widehat{\theta }$ will be
considered optimal when it minimizes the relevant risk, no matter what the
true state of nature $\theta ^{\ast }$ happens to be. This constitutes a key
caveat that is often ignored in discussions of these approaches. When viewed
as a game against Nature, the decision maker selects action $a\mathit{\ }$%
from $A,$ irrespective of what value $\theta ^{\ast }$ Nature has chosen.
That is, $\theta ^{\ast }$ plays no role in selecting the optimal rules
since the latter have nothing to do with the true value $\theta ^{\ast }$ of 
$\theta .$ To avoid any misreading of this line of reasoning, it is
important to emphasize that `the true value $\theta ^{\ast }$' is shorthand
for saying that `data $\mathbf{x}_{0}$ constitute a typical realization of
the sample $\mathbf{X}$ with distribution $f(\mathbf{x};\mathit{\theta }%
^{\ast })$'; see Spanos and Mayo (2015). This should be contrasted with the
notion of optimality in frequentist inference that gives $\theta ^{\ast }$
center stage, in the sense that it evaluates the capacity of the inference
procedure to inform the modeler about $\theta ^{\ast }\mathbf{;}$ no other
value is relevant. According to Reid (2015):

\textsf{\textquotedblleft A statistical model is a family of probability
distributions [}$\mathcal{M}_{\mathbf{\theta }}(\mathbf{x})$\textsf{], the
central problem of statistical inference being to identify which member of
the family [}$\theta ^{\ast }$\textsf{] generated the data of
interest.\textquotedblright\ (p. 418)}\vspace*{-0.15in}

\section{The Bayesian approach\protect\vspace*{-0.06in}}

To shed further light on the affinity between the decision-theoretic
framework and the Bayesian approach, let us take a closer look at the latter.%
\vspace*{-0.15in}

\subsection{\textbf{Bayesian inference and} its primary objective\protect%
\vspace*{-0.06in}}

A key argument in favor of the Bayesian approach is often its simplicity in
the sense that all forms of inference revolve around a single function, the
posterior distribution: $\pi(\mathbf{\theta\mathit{\mid}x}_{0})\mathit{%
\propto }\pi(\mathbf{\theta})\mathit{\cdot}f(\mathbf{x}_{0}\mathit{\mid}%
\mathbf{\theta}),\ \forall\mathbf{\theta}\mathit{\in}\Theta.$ This, however,
is only half the story. The other half is how the posterior distribution is
utilized to yield `optimal' inferences. The issue of optimality, however, is
intrinsically related to what the primary objective of Bayesian inference is.

An outsider looking at Bayesian approach would surmise that its primary
objective is to yield `the probabilistic ranking' (ordering) of all values
of $\mathbf{\theta}\mathit{\ }$in $\Theta.$ The modeling begins with an a
priori probabilistic ranking based on $\pi(\mathbf{\theta}),\ \forall\mathbf{%
\theta }\mathit{\in}\Theta,$ which is revised after observing $\mathbf{x}%
_{0} $ to derive $\pi(\mathbf{\theta\mathit{\mid}x}_{0}),\ \forall\mathbf{%
\theta }\mathit{\in}\Theta$; hence the key role of the quantifier $\forall 
\mathbf{\theta}\mathit{\in}\Theta$. Indeed, O'Hagan's (1994) argues that the
revised probabilistic ranking \textit{is} the inference:

\textsf{\textquotedblleft Having obtained the posterior density }$\pi(%
\mathbf{\theta}\mathit{\mid}\mathbf{x}_{0})$\textsf{, the final step of the
Bayesian method is to derive from it suitable inference statements. The most
usual inference question is this: After seeing the data }$\mathbf{x}_{0} $%
\textsf{, what do we now know about the parameter }$\mathbf{\theta.}$\textsf{%
\ The only answer to this question is to present the entire posterior
distribution.\textquotedblright\ (p. 6)}

In light of that, the question that naturally arises is: what does the
revised probabilistic ranking, based on $\pi(\mathbf{\theta\mathit{\mid}x}%
_{0}),\ \forall\mathbf{\theta}\mathit{\in}\Theta,$ convey about the
underlying data generating mechanism in (\ref{smo}), assumed to have given
rise to data $\mathbf{x}_{0}$? That is, what does this ranking have to do
with learning about the `true' value $\mathbf{\theta}^{\ast}?$ It is not
obvious why the highest ranked value $\widetilde{\mathbf{\theta}}(\mathbf{x}%
_{0})$ (mode) or some other feature of the posterior distribution pinpoints $%
\mathbf{\theta }^{\ast}$ better than any other value. As mentioned above, $%
\mathbf{\theta }^{\ast}$ plays no role in selecting the optimal rule, since
the latter revolves exclusively around the relevant risk. In contrast,
learning from data $\mathbf{x}_{0}$ about the unique value $\mathbf{\theta}%
^{\ast}$ makes perfectly good sense in frequentist inference since $\mathbf{%
\theta}$ is viewed as an unknown \textit{constant} that gave rise to $%
\mathbf{x}_{0}$. This issue highlights the key built-in tension between the
frequentist and Bayesian approaches.

If the primary objective of Bayesian inference is not the revised
probabilistic ranking of all $\mathbf{\theta}\mathit{\in}\Theta,\ $what is
it? The answer is that the decision-theoretic perspective came in to refocus
and append to Bayesian inference. O'Hagan, echoing earlier views by Lindley
(1965) and Tiao and Box (1975) in contrasting frequentist (classical)
inferences with Bayesian inferences, argues:

\textsf{\textquotedblleft Classical inference theory is very concerned with
constructing \textit{good} inference rules. The primary concern of Bayesian
inference, ..., is entirely different. The objective is to extract
information concerning }$\mathbf{\theta}$\textsf{\ from the posterior
distribution, and to present it helpfully via effective summaries. There are
two criteria in this process. The first is to identify interesting features
of the posterior distribution. ... The second criterion is good
communication. Summaries should be chosen to convey clearly and succinctly
all the features of interest. ... In Bayesian terms, therefore, a good
inference is one which contributes effectively to appropriating the
information about} $\mathbf{\theta}$\textsf{\ which is conveyed by the
posterior distribution.\textquotedblright\ (p. 14)}

Clearly, O'Hagan's attempt to define what is a `good' Bayesian inference
begs the question: what does constitute `effective appropriation of
information about $\mathbf{\theta}$' mean, apart from the probabilistic
ranking? Putting that aside, in his attempt to defend his stance that the
entire posterior distribution is the inference, O'Hagan argues that criteria
for `optimal' Bayesian inferences are only \textit{parasitical} on the
Bayesian approach and enter the picture via the decision theoretic
perspective:

\textsf{\textquotedblleft... a study of decision theory has two potential
benefits. First, it provides a link to classical inference. It thereby shows
to what extent classical estimators, confidence intervals and hypotheses
tests can be given a Bayesian interpretation or motivation. Second, it helps
identify suitable summaries to give Bayesian answers to stylized inference
questions which classical theory addresses.\textquotedblright\ (p. 14)}

Both of the above mentioned potential benefits to the Bayesian approach, are
misleading for two reasons. First, the link\ between the decision-theoretic
and the classical (frequentist) inference is fraught with misleading
definitions and unclarities with respect to the reasoning and objectives of
the latter. For instance, the quantifier `$\forall\theta\mathit{\in}\Theta$'
used to define `optimal' estimators with respect to particular loss
functions is at odds with frequentist inference. Second, the claim
concerning Bayesian answers to frequentist questions of interest is
misplaced because the former provides no real answers to the frequentist
primary question of interest which pertains to learning about $%
\theta^{\ast}. $ A Bayes rule offers very little, if anything, relevant in
learning what the value $\theta^{\ast}$ that gave rise to $\mathbf{x}_{0}$
is. Let us unpack this answer in more detail.

Substituting the risk function in (\ref{rf}) into the Bayes risk in (\ref{ri}%
), one can show that:$\vspace*{-0.06in}$%
\begin{equation}
\begin{array}{cl}
R_{B}(\widehat{\theta }) & \mathit{=}\int_{\theta \in \Theta }\left( \int_{%
\mathbf{x}\in \mathbb{R}_{X}^{n}}L(\theta ,\widehat{\theta }(\mathbf{x}))f(%
\mathbf{x};\theta )d\mathbf{x}\right) \pi (\theta )d\theta \mathit{=} \\ 
& \mathit{=}\int_{\mathbf{x}\mathit{\in }\mathbb{R}_{X}^{n}}\int_{\theta \in
\Theta }L(\theta ,\widehat{\theta }(\mathbf{x)})f(\mathbf{x}\mathit{\mid }%
\theta )\pi (\theta )d\theta d\mathbf{x}\mathit{=} \\ 
& \mathit{=}\int_{\mathbf{x}\mathit{\in }\mathbb{R}_{X}^{n}}\left\{
\int_{\theta \in \Theta }L(\theta ,\widehat{\theta }(\mathbf{x)})\pi (\theta 
\mathit{\mid }\mathbf{x})d\theta \right\} m(\mathbf{x})d\mathbf{x,}%
\end{array}%
\vspace*{-0.06in}  \label{br}
\end{equation}%
where $m(\mathbf{x})\mathit{=}\int_{\theta \in \Theta }f(\mathbf{x};\theta
)d\theta $; see Bansal (2007). The second and third equalities presume that
one can reverse the order of integration (a technical issue), and treat $f(%
\mathbf{x};\theta )$ as the joint distribution of $\mathbf{X}\ $and $\theta $
so that:$\vspace*{-0.06in}$ 
\[
\begin{array}{c}
f(\mathbf{x};\theta )\mathit{=}f(\mathbf{x}\mathit{\mid }\theta )\pi (\theta
)\mathit{=}\pi (\theta \mathit{\mid }\mathbf{x})m(\mathbf{x}).%
\end{array}%
\vspace*{-0.06in}
\]%
$\ $The latter raises a number of questionable hand-waving moves because it
muddies the distinction between $\mathbf{x},$ a generic value of $\mathbb{R}%
_{X}^{n},$ and the particular value $\mathbf{x}_{0}$; see Spanos (2015). In
light of (\ref{br}), a Bayesian estimate is `optimal' relative to a
particular loss function $L(\widehat{\theta }(\mathbf{X}),\theta ),$ when it
minimizes $R_{B}(\widehat{\theta })$. This makes it clear that what
constitutes an `optimal' Bayesian estimate is primarily determined by $L(%
\widehat{\theta }(\mathbf{X}),\theta )$ (Schervish, 1995):\medskip \newline
(i) when $L_{2}(\widehat{\theta },\theta )\mathbf{=}(\widehat{\theta }%
-\theta )^{2}$ the Bayes estimate $\widehat{\theta }$ is the \textit{mean}
of $\pi (\theta \mathit{\mid }\mathbf{x}_{0}),$\medskip \newline
(ii) when $L_{1}(\widetilde{\theta },\theta )\mathbf{=}|\widetilde{\theta }%
-\theta |$ the Bayes estimate $\widetilde{\theta }$ is the \textit{median}
of $\pi (\theta \mathit{\mid }\mathbf{x}_{0}),$\medskip \newline
(iii) when $L_{0-1}(\overline{\theta },\theta )\mathbf{=}\delta (\overline{%
\theta },\theta )\mathbf{=}\left\{ 
\begin{array}{cc}
0 & \text{for }\left\vert \overline{\theta }\mathbf{-}\theta \right\vert
<\varepsilon  \\ 
1 & \text{for }\left\vert \overline{\theta }\mathbf{-}\theta \right\vert
\geq \varepsilon 
\end{array}%
\right. ,$ for $\varepsilon \mathit{>}0,$ the Bayes estimate $\overline{%
\theta }$ is the \textit{mode} of $\pi (\theta \mathit{\mid }\mathbf{x}_{0})$%
.\medskip 

In practice, the most widely used loss function is the square:$\vspace*{%
-0.05in}$%
\[
\begin{array}{c}
L_{2}(\widehat{\theta }(\mathbf{X});\theta )\mathbf{=}(\widehat{\theta }(%
\mathbf{X})-\theta )^{2},\ \forall \theta \mathit{\in }\Theta ,%
\end{array}%
\]%
whose risk function is the decision-theoretic \textit{Mean Square Error (MSE}%
$_{1}$\textit{)}:$\vspace*{-0.05in}$%
\begin{equation}
\begin{array}{c}
R(\theta ,\widehat{\theta })\mathbf{=}E(\widehat{\theta }(\mathbf{X})\mathbf{%
-}\theta )^{2}\mathbf{=}\text{\textsf{MSE}}_{1}(\widehat{\theta }(\mathbf{X}%
);\theta ),\ \forall \theta \mathit{\in }\Theta .%
\end{array}
\label{MSE}
\end{equation}%
Surprising, however, this definition of the MSE, denoted by MSE$_{1},$ is
different from the frequentist MSE, which is defined by:$\vspace*{-0.05in}$%
\begin{equation}
\begin{array}{c}
MSE(\widehat{\theta }_{n}(\mathbf{X});\theta ^{\ast })\mathbf{=}E(\widehat{%
\theta }_{n}(\mathbf{X})\mathit{-}\theta ^{\ast })^{2}.%
\end{array}%
\vspace*{-0.05in}  \label{mse0}
\end{equation}%
The key difference is that (\ref{mse0}) is defined at the point $\theta 
\mathit{=}\theta ^{\ast },$ as opposed to $\forall \theta \mathit{\in }%
\Theta $. Unfortunately, statistics textbooks adopt one of the two
definitions of the MSE -- either at $\theta \mathbf{=}\theta ^{\ast }$ or $%
\forall \theta \mathit{\in }\Theta $ -- and ignore (or seem unaware) of the
other. At first sight, his difference might appear pedantic, but it turns
out that it has very serious implications for the relevant theory of
optimality for the frequentist vs. Bayesian inference procedures. Indeed,
reliance on $\forall \theta \mathit{\in }\Theta $ undermines completely the
relevance of admissibility in frequentist inference.

\textbf{Admissibility}. An estimator $\widetilde{\theta }(\mathbf{X})$ is 
\textit{inadmissible} if there exists another estimator $\widehat{\theta }(%
\mathbf{X})$ such that:$\vspace*{-0.05in}$%
\begin{equation}
\begin{array}{c}
R(\theta ,\widehat{\theta })\leq R(\theta ,\widetilde{\theta }),\text{ }%
\forall \theta \mathit{\in }\Theta ,%
\end{array}%
\vspace*{-0.06in}  \label{a}
\end{equation}%
and the strict inequality ($<$) holds for at least one value of $\theta .$
Otherwise, $\widetilde{\theta }(\mathbf{X})$ is said to be \textit{admissible%
} with respect to the loss function $L(\theta ,\widehat{\theta }).$\vspace*{%
-0.15in}

\subsection{The duality between loss functions and priors\protect\vspace*{%
-0.06in}}

The built-in affinity between the decision-theoretic set-up and Bayesian
inference is cemented by a \textit{duality} result between loss functions
and prior distributions; see Robert (2001). This duality stems from the fact
that minimizing the Bayes risk:$\vspace*{-0.1in}$%
\[
\begin{array}{c}
R_{B}(\widehat{\theta })\mathit{=}\int_{\theta \in \Theta }R(\widehat{\theta 
},\theta )\pi (\theta )d\theta ,%
\end{array}%
\vspace*{-0.1in}
\]
is equivalent to minimizing the integral:$\vspace*{-0.1in}$%
\[
\begin{array}{c}
\int_{\theta \in \Theta }L(\widehat{\theta }(\mathbf{X}),\theta )\pi (\theta 
\mathit{\mid }\mathbf{x})d\theta .%
\end{array}%
\vspace*{-0.1in}
\]
This result brings out two important features of Bayesian inference. First,
it confirms the minor role played by the quantifier $\mathbf{x}\mathit{\in }%
\mathbb{R}_{X}^{n}$ in both Bayesian and decision-theoretic optimality
theory of inference. Second, it\ indicates that $L(\theta ,\widehat{\theta }%
) $ and $\pi (\theta )$ are perfect substitutes with respect to any weight
function $w(\theta )\mathit{>}0,$ $\forall \theta \mathit{\in }\Theta $, in
the derivation of Bayes rules. Modifying the loss function or the prior
yields the same result:

\textsf{\textquotedblleft... the problem of estimating }$\theta$\textsf{\
with a modified (weighted) loss function is identical to the problem with a
simple loss but with modified hyperparameters of the prior distribution
while the form of the prior distribution does not change.\textquotedblright\
(Srivastava et al., 2014, p. 522)}

This implies that in practice a Bayesian could derive a particular Bayes
estimate by attaching the weight to the loss function or to the prior
distribution depending on which derivation is easier; see Bansal (2007),
Srivastava et al. (2014). \vspace*{-0.12in}

\subsection{\textbf{Bayes rule and admissibility}\protect\vspace*{-0.06in}}

As argued in the sequel, it should come as no surprise to learn that Bayes
rules dominate all other rules when admissibility is given center stage.

A Bayes rule $\widehat{\theta}_{B}(\mathbf{X})$ with respect to a prior
distribution $\pi(\theta)$ is:

(i) \textit{Admissible}, under certain regularity conditions, including when 
$\widehat{\theta}_{B}(\mathbf{X})$ is unique up to equivalence relative to
the same risk function.

(ii) \textit{Minimax} when $R(\theta,\widehat{\theta}_{B})\mathit{=}c\mathit{%
<}\infty.$

(iii) An admissible estimate $\widehat{\theta}(\mathbf{X})$ is either Bayes $%
\widehat{\theta}_{B}(\mathbf{X})$ or the limit of a sequence of Bayes rules;
see Wasserman (2004), Srivastava et al. (2014).

These results have been used as evidence for the superiority of the Bayesian
perspective, and led to the intimation that an effective way to generate
optimal frequentist procedures is to find the Bayes solution using a
reasonable prior and then examine their frequentist properties to see
whether it is satisfactory from the latter viewpoint; see Rubin (1984),
Gelman et al (2004).

Even if one were to agree that Bayes rules and admissible estimators largely
coincide, the importance of such a result hinges on the appropriateness of
admissibility for frequentist estimators. \vspace*{-0.15in}

\section{Loss functions and admissibility revisited\protect\vspace*{-0.06in}}

The claim to be discussed in this section is that the notions of a `loss
function' and `admissibility' are incompatible with the optimal theory of
frequentist estimation largely framed by Fisher; see Savage (1976).\vspace*{%
-0.15in}

\subsection{Admissibility as a minimal property\protect\vspace*{-0.06in}}

The following example is used to call into question the notion of a loss
function and the associated property of admissibility for optimal
frequentist estimators.

\textbf{Example}. In the context of the simple Normal model:\vspace*{-0.08in}%
\begin{equation}
\begin{array}{c}
X_{k}\backsim\text{\textsf{NIID}}(\theta,1),\ k\mathbf{=}1,2,...,n,\ \text{%
for }n>2,%
\end{array}
\vspace*{-0.08in}  \label{snm}
\end{equation}
let us use the decision-theoretic notion of MSE$_{1}$ in (\ref{MSE}) to
compare two estimators of $\theta$:%
\[
\begin{tabular}{l}
(i) the Maximum Likelihood Estimator (MLE): $\overline{X}_{n}\mathbf{=}\frac{%
1}{n}\sum\nolimits_{k=1}^{n}X_{k},\medskip$ \\ 
$\text{(ii) the `crystalball' estimator: }\theta_{cb}\mathbf{=}7405926,\text{
}\forall\mathbf{x}\mathit{\in}\mathbb{R}_{X}^{n}.$%
\end{tabular}%
\]
When compared on admissibility grounds, both estimators are admissible, and
thus equally acceptable. Common sense, however, suggests that if a
particular criterion of optimality cannot distinguish between $\overline{X}%
_{n}$ [a strongly consistent, unbiased, fully efficient and sufficient
estimator] and $\theta_{cb},$ an arbitrarily chosen real number that ignores
the data altogether, is not much of a minimal property.

A moment's reflection suggests that the inappropriateness of admissibility
stems from its reliance on the quantifier `$\forall\theta\mathit{\in}\Theta$%
'. The admissibility of $\theta_{cb}$ arises from the fact that for certain
values of $\theta$ close to $\theta_{cb}$, say $\theta\mathit{\in}(\theta
_{cb}\mathbf{\pm}\frac{\lambda}{\sqrt{n}}),$\ for $0\mathit{<}\lambda 
\mathit{<}1,\ \theta_{cb}$ is `better' than $\overline{X}_{n}$ on MSE$_{1}$
grounds:\vspace*{-0.08in}%
\begin{equation}
\begin{array}{c}
MSE_{1}(\overline{X}_{n};\theta)\mathbf{=}\frac{1}{n}>MSE_{1}(\theta
_{cb};\theta)\leq\frac{\lambda^{2}}{n}\text{ for }\theta\mathit{\in}%
(\theta_{cb}\mathbf{\pm}\frac{\lambda}{\sqrt{n}}).%
\end{array}
\vspace*{-0.08in}  \label{lt}
\end{equation}
Given that the primary objective of a frequentist estimator is to pin-point $%
\theta^{\ast},$ the result in (\ref{lt}) seems totally irrelevant as a gauge
of its capacity to achieve that!

This example indicates that admissibility is totally ineffective as a 
\textit{minimal }property because it does not filter out $\theta_{cb},$ the
worst possible estimator! Instead, it excludes potentially good estimators
like the \textit{sample median}; see Cox and Hinkley (1974). This highlights
the `extreme relativism' of admissibility to the particular loss function, $%
L_{2}(\widehat{\theta}(\mathbf{X});\theta)$, in this case. For the absolute
loss function $L_{1}(\widehat{\theta}(\mathbf{X});\theta)\mathbf{=}|\widehat{%
\theta}(\mathbf{X})-\theta|$, however, the sample median would have been the
optimal estimator. What determines which loss function is appropriate in
particular cases?

Despite his wholehearted embrace of the decision-theoretic framing, Lehmann
(1984) warned statisticians about the perils of arbitrary loss functions:

\textsf{\textquotedblleft It is argued that the choice of a loss function,
while less crucial than that of the model, exerts an important influence on
the nature of the solution of a statistical decision problem, and that an
arbitrary choice such as squared error may be baldly misleading as to the
relative desirability of the competing procedures.\textquotedblright\ (p.
425)}

A strong case can be made that \textit{the} key minimal property (necessary
but not sufficient) for frequentist estimation is \textit{consistency}, an
extension of the Law of Large Numbers to estimators. For instance,
consistency would have eliminated both $\widetilde{\theta}$ and $\theta_{cb}$
from consideration; they are both inconsistent. This makes intuitive sense
because if an estimator $\widehat{\theta}(\mathbf{X})$ cannot pinpoint $%
\theta^{\ast}$ with an infinite data information, it should be considered
impertinent. In contrast, there is nothing in the notion of admissibility
that advances learning from data about $\theta^{\ast}$.

Further to relative (to particular loss functions) efficiency being a
dubious property for frequentist estimators, the pertinent measure of finite
sample precision for frequentist estimators is full efficiency, which is
defined relative to the assumed statistical model (\ref{smo}), and not some
arbitrary loss function based on information other than the data.\vspace*{%
-0.12in}

\subsection{\textbf{Stein's paradox and admissibility}\protect\vspace*{%
-0.06in}}

The quintessential example that has bolstered the appeal of the Bayesian
claims concerning admissibility is the James-Stein estimator (Efron and
Morris, 1973), that gave rise to an extensive literature on \textit{%
shrinkage estimators}; see Saleh (2006).

Let $\mathbf{X}\mathit{:=}(X_{1},X_{2},...,X_{m})$ be independent sample
from a Normal distribution:\vspace*{-0.06in}%
\begin{equation}
\begin{array}{c}
X_{k}\backsim \text{\textsf{NI}}(\theta _{k},\sigma ^{2}),\ k\mathbf{=}%
1,2,...,m,%
\end{array}%
\vspace*{-0.06in}  \label{sn}
\end{equation}%
where $\sigma ^{2}$ is known. Using the notation $\mathbf{\theta }$:=$%
(\theta _{1},\theta _{2},...,\theta _{m})$ and $\mathbf{I}_{m}$:=diag($%
1,1,...,1$), this can be denoted by:\vspace*{-0.06in}%
\[
\begin{array}{c}
\mathbf{X}\backsim \text{\textsf{N}}(\mathbf{\theta },\sigma ^{2}\mathbf{I}%
_{m}).%
\end{array}%
\vspace*{-0.06in}
\]%
Find an optimal estimator $\widehat{\mathbf{\theta }}(\mathbf{X})$ of $%
\mathbf{\theta }$ with respect to the square `overall' loss function:%
\begin{equation}
\begin{array}{c}
L_{2}(\mathbf{\theta },\widehat{\mathbf{\theta }}(\mathbf{X}))\mathbf{=}(%
\mathbf{\parallel }\widehat{\mathbf{\theta }}(\mathbf{X})\mathit{-}\mathbf{%
\theta }\mathit{\parallel }^{2})\mathit{=}\sum_{k=1}^{m}(\widehat{\theta }%
_{k}(\mathbf{X})\mathit{-}\theta _{k})^{2}.%
\end{array}
\label{lm}
\end{equation}%
Stein (1956) astounded the statistical world by showing that for $m\mathbf{=}%
2$ the Least-Squares (LS) estimator $\widehat{\mathbf{\theta }}_{LS}(\mathbf{%
X})\mathit{=}\mathbf{X}$ is admissible, but for $m>2$ $\widehat{\mathbf{%
\theta }}_{LS}(\mathbf{X})$ is inadmissible. Indeed, James and Stein (1961)
were able to come up with a nonlinear estimator:$\vspace*{-0.06in}$%
\begin{equation}
\begin{array}{c}
\widehat{\mathbf{\theta }}_{JS}(\mathbf{X})\mathbf{=}\left( 1-\frac{%
(m-2)\sigma ^{2}}{\parallel \mathbf{X}\parallel ^{2}}\right) \mathbf{X,}%
\end{array}%
\vspace*{-0.06in}  \label{js}
\end{equation}%
that became known as the James-Stein estimator, which dominates $\widehat{%
\mathbf{\theta }}_{LS}(\mathbf{X})\mathit{=}\mathbf{X}$ in MSE$_{1}$ terms
by demonstrating that:\vspace*{-0.12in}%
\begin{equation}
\begin{array}{c}
\text{\textsf{MSE}}_{1}\text{(}\widehat{\mathbf{\theta }}_{JS}(\mathbf{X});%
\mathbf{\theta }\text{)}<\text{\textsf{MSE}}_{1}\text{(}\widehat{\mathbf{%
\theta }}_{LS}(\mathbf{X});\mathbf{\theta }\text{), }\forall \mathbf{\theta }%
\mathit{\in }\mathbb{R}^{m}.%
\end{array}%
\vspace*{-0.06in}  \label{in}
\end{equation}%
It turns out that $\widehat{\mathbf{\theta }}_{JM}(\mathbf{X})$ is also
inadmissible for $m>2$ and dominated by the modified James-Stein estimator
that is \textit{admissible}:\vspace*{-0.06in}%
\[
\begin{array}{c}
\widehat{\mathbf{\theta }}_{JS}^{+}(\mathbf{X})\mathbf{=}\left( 1-\frac{%
(m-2)\sigma ^{2}}{\parallel \mathbf{X}\parallel ^{2}}\right) ^{+}\mathbf{X,}%
\end{array}%
\vspace*{-0.06in}
\]%
where $\left( z\right) ^{+}\mathit{=}\max (0,z);$ see Wasserman (2004).

The traditional interpretation of this result is that for the Normal,
Independent model in (\ref{sn}), the James--Stein estimator (\ref{js}) of $%
\mathbf{\theta}$:=$(\theta_{1},\theta_{2},...,\theta_{m}),$ for $m>2,$
reduces the \textit{overall} MSE$_{1}$ in (\ref{lm}). This result seems to
imply that one will `do better' (in overall MSE$_{1}$ terms) by using a
combined nonlinear (shrinkage) estimator, instead of estimating these means
separately. What is surprising about this result is that there is no
statistical reason (due to independence) to connect the inferences
pertaining to the different individual means, and yet the obvious estimator
(LS) is inadmissible.

As argued next, this result calls into question the appropriateness of the
notion of admissibility with respect to a particular loss function, and not
the judiciousness of frequentist estimation.\vspace*{-0.15in}

\section{Frequentist inference and learning from data\protect\vspace*{-0.06in%
}}

The objectives and underlying reasoning of frequentist inference are
inadequately discussed in the statistics literature. As a result some of its
key differences with Bayesian inference remain beclouded.\vspace*{-0.15in}

\subsection{Frequentist approach: primary objective and reasoning\protect%
\vspace*{-0.06in}}

All forms of parametric frequentist inference begin with a prespecified
statistical model $\mathcal{M}_{\mathbf{\theta}}(\mathbf{x})\mathbf{=}\{f(%
\mathbf{x};\mathbf{\theta}),\ \mathbf{\theta}\mathit{\in}\Theta \},\ \mathbf{%
x}\mathit{\in}\mathbb{R}_{X}^{n}.$ This model is chosen from the set of all
possible models that could have given rise to data $\mathbf{x}_{0} $\textbf{:%
}$\mathbf{=}(x_{1},...,x_{n}),$ by selecting the probabilistic structure for
the underlying stochastic process $\{X_{t},\ t\mathit{\in }\mathbb{N}\mathit{%
:=}(1,2,...,n,...)\}$ in such a way so as to render the observed data $%
\mathbf{x}_{0}$ a `typical' realization thereof. In light of the fact that
each value of $\mathbf{\theta}\mathit{\in}\Theta$ represents a different
element of the family of models represented by $\mathcal{M}_{\mathbf{\theta}%
}(\mathbf{x}),$ the primary objective of frequentist inference is to learn
from data about the `true' model:\vspace*{-0.06in}%
\begin{equation}
\begin{array}{c}
\mathcal{M}^{\ast}(\mathbf{x})\mathbf{=}\{f(\mathbf{x};\mathbf{\theta}^{\ast
})\},\ \mathbf{x}\mathit{\in}\mathbb{R}_{X}^{n},%
\end{array}
\vspace*{-0.06in}  \label{dgm}
\end{equation}
where $\mathbf{\theta}^{\ast}$ denotes the true value of $\mathbf{\theta}$
in $\Theta$. The `typicality' is testable vis-a-vis the data $\mathbf{x}_{0} 
$ using misspecification testing; see Spanos(2006).

The frequentist approach relies on two modes of reasoning for inference
purposes:$\ $%
\[
\begin{tabular}{ll}
\textit{Factual} (estimation, prediction): & $f(\mathbf{x};\theta^{\ast }),\
\forall\mathbf{x}\mathit{\in}\mathbb{R}_{X}^{n},\medskip$ \\ 
\textit{Hypothetical} (hypothesis testing): & $f(\mathbf{x};\theta_{0}),$ $f(%
\mathbf{x};\theta_{1}),\ \forall\mathbf{x}\mathit{\in}\mathbb{R}_{X}^{n},$%
\end{tabular}
\ 
\]
where $\theta^{\ast}$ denotes the true value of $\theta$ in $\Theta$, and $%
\theta_{i},\ i\mathit{=}0,1$ denote hypothesized values of $\theta$
associated with the hypotheses, $H_{0}$: $\theta_{0}\mathit{\in}\Theta_{0} $%
, $H_{1}$: $\theta_{1}\mathit{\in}\Theta_{1},$ where $\Theta_{0}$ and $%
\Theta_{1}$ constitute a partition of $\Theta.$

A frequentist estimator $\widehat{\theta}$ aims to pin-point $\theta^{\ast}$%
, and its optimality is evaluated by how effectively it achieves that.
Similarly, a test statistic usually compares a good estimator $\widehat{%
\theta }$ of $\theta$ with a prespecified value $\theta_{0},$ but behind $%
\widehat{\theta}$ is the value $\theta^{\ast}$ assumed to have generated
data $\mathbf{x}_{0}.$ Hence, the hypothetical reasoning is used in testing
to learn about $\theta^{\ast},$ and has nothing to do with all possible
values of $\theta$ in $\Theta.$

This contradicts misleading claims by Bayesian textbooks (Robert, 2001, p.
61):

\textsf{\textquotedblleft The frequentist paradigm relies on this criterion
[risk function] to compare estimators and, if possible, to select the best
estimator, the reasoning being that estimators are evaluated on their
long-run performance for all possible values of the parameter }$\theta.$%
\textsf{\textquotedblright}

\hspace*{-0.25in}Contrary to this claim, the only relevant value of $\theta$
in evaluating the `optimality' of $\widehat{\theta}$ is $\theta^{\ast}.$
Such misleading claims stem from an apparent confusion between the
existential and universal quantifiers in framing certain inferential
assertions.

The existence of $\theta^{\ast}$ can be formally defined using the \textit{%
existential} quantifier:$\vspace*{-0.06in}$ 
\[
\begin{tabular}{ll}
$\exists\theta^{\ast}\mathit{\in}\Theta:$ & there exists a $\theta^{\ast }%
\mathit{\in}\Theta$ such that.%
\end{tabular}
\ \vspace*{-0.06in}
\]
This introduces a potential conflict between the \textit{existential} and
the \textit{universal} quantifier `$\forall\theta\mathit{\in}\Theta$'
because neither the decision theoretic nor the Bayesian approach explicitly
invoke $\theta^{\ast}$. Decision-theoretic and Bayesian rules are considered
optimal when they minimize the expected loss $\forall\theta\mathit{\in}%
\Theta,$ no matter what $\theta^{\ast}$ happens to be. How different the two
quantifiers are, can be demonstrated using elementary logic. The logical
connective for negation ($\lnot$) can be used to define the following
equivalence relationships between the two quantifiers:$\vspace*{-0.08in}$%
\[
\begin{tabular}{ll}
(i) $\exists\theta^{\ast}\mathit{\in}\Theta\iff\lnot\forall\mathbf{\theta }%
\mathit{\notin}\Theta,$ & (ii) $\forall\mathbf{\theta}\mathit{\in}\Theta
\iff\lnot\exists\theta^{\ast}\mathit{\notin}\Theta$%
\end{tabular}
\vspace*{-0.08in}
\]
Given that (i)-(ii) involve double negations, the two quantifiers could not
be further apart.

At first sight the quantifier $\forall \theta \mathit{\in }\Theta $ seems
rather innocuous and natural in the context of statistical inference. It
seems intuitively obvious that one should care about the behavior of an
estimator $\widehat{\theta }$ for all possible values of $\theta .$ This is
misleading intuition, however, since the behavior of $\widehat{\theta },$
for all $\theta \mathit{\in }\Theta $, although relevant, is not what
determines how effective a frequentist estimator is at pin-pointing $\theta
^{\ast };$ what matters is its behavior around $\theta ^{\ast }$. Assessing
its effectiveness calls for evaluating (deductively) the sampling
distribution of $\widehat{\theta }$ under $\theta \mathit{=}\theta ^{\ast },$
or hypothetical values $\theta _{0}$ and $\theta _{1},$ and not for all
possible values of $\theta $ in $\Theta .$ Let's unpack the details of this
claim.\vspace*{-0.15in}

\subsection{Frequentist estimation\protect\vspace*{-0.06in}}

The underlying reasoning for frequentist estimation is \textit{factual}%
\textbf{,} in the sense the optimality of an estimator is appraised in terms
of its generic capacity of $\widehat{\theta }_{n}(\mathbf{X})$ to\ \textit{%
zero-in} on (pinpoint) the true value $\theta ^{\ast }$, whatever the sample
realization $\mathbf{X\mathit{=}x}_{0}$. Optimal properties like
consistency, unbiasedness, full efficiency, sufficiency, etc., calibrate
this generic capacity using its sampling distribution of $\widehat{\theta }%
_{n}(\mathbf{X})$ evaluated under $\theta \mathbf{=}\theta ^{\ast }$i.e., in
terms of $f(\widehat{\theta }_{n}(\mathbf{x});\theta ^{\ast }),$ for $%
\mathbf{x}\mathit{\in }\mathbb{R}_{X}^{n}.$ For instance, \textit{strong
consistency} asserts that as $n\rightarrow \infty ,$ $\widehat{\theta }_{n}(%
\mathbf{X})$ will zero-in on $\theta ^{\ast }$ almost surely:$\vspace*{%
-0.06in}$%
\[
\begin{array}{c}
\mathbb{P}(\underset{n\rightarrow \infty }{\lim }\widehat{\theta }_{n}(%
\mathbf{X})\mathbf{=}\theta ^{\ast })\mathbf{=}1.%
\end{array}%
\]%
Similarly, \textit{unbiasedness} asserts that the sampling distribution of $%
\widehat{\theta }_{n}(\mathbf{X})$ has a mean equal to $\theta ^{\ast }:$%
\vspace*{-0.06in}%
\[
\begin{array}{c}
E(\widehat{\theta }_{n}(\mathbf{X}))\mathbf{=}\theta ^{\ast }.%
\end{array}%
\]%
In this sense both of these optimal properties are defined at the point $%
\theta $=$\theta ^{\ast }$. This is achieved by using \textit{factual
reasoning}, i.e. evaluating the sampling distribution of $\widehat{\theta }%
_{n}(\mathbf{X})$ under the true state of nature ($\theta $=$\theta ^{\ast }$%
), without having to know $\theta ^{\ast }.$ This is in contrast to using
loss functions, such as (\ref{w}), which are defined in terms of $\theta
^{\ast }$ but are rendered non-operational without knowing $\theta ^{\ast }$.

\textbf{Example}. In the case of the simple Normal model in (\ref{snm}) the
point estimator, $\overline{X}_{n}$ is consistent, unbiased, fully
efficient, sufficient, with a sampling distribution:\vspace*{-0.05in}%
\begin{equation}
\begin{array}{c}
\overline{X}_{n}\backsim \text{\textsf{N}}(\theta ,\frac{1}{n}).%
\end{array}
\label{sm}
\end{equation}%
What is not usually explicitly stated is that the evaluation of that
distribution is \textit{factual}, i.e. $\theta \mathbf{=}\theta ^{\ast }$,
and should formally denoted by:\vspace*{-0.05in}%
\[
\begin{array}{c}
\overline{X}_{n}\overset{\theta \mathbf{=}\theta ^{\ast }}{\backsim }\mathsf{%
N}(\theta ^{\ast },\frac{1}{n}).%
\end{array}%
\]%
When $\overline{X}_{n}$ is standardized, it yields the pivotal function:%
\vspace*{-0.07in}%
\begin{equation}
\begin{array}{c}
d(\mathbf{X;}\theta )\mathit{:=}\sqrt{n}\left( \overline{X}_{n}-\theta
^{\ast }\right) \overset{\theta =\theta ^{\ast }}{\backsim }\text{\textsf{N}}%
(0,1),%
\end{array}
\label{pq}
\end{equation}%
whose distribution only holds for the true $\theta ^{\ast },$ and no other
value. This provides the basis for constructing a $(1\mathbf{-}\alpha )$
Confidence Interval (CI):\vspace*{-0.1in}%
\begin{equation}
\begin{array}{cc}
\mathbb{P}\left( \overline{X}_{m}-c_{\frac{\alpha }{2}}(\frac{1}{\sqrt{n}}%
)\leq \theta \leq \overline{X}_{n}+c_{\frac{\alpha }{2}}(\frac{1}{\sqrt{n}}%
);\theta \mathbf{=}\theta ^{\ast }\right) \mathbf{=}1\mathbf{-}\alpha , & 
\end{array}%
\vspace*{-0.1in}  \label{ci}
\end{equation}%
which asserts that the random interval $[\overline{X}_{n}\mathbf{-}c_{\frac{%
\alpha }{2}}(\frac{s}{\sqrt{n}}),\ \overline{X}_{n}\mathbf{+}c_{\frac{\alpha 
}{2}}(\frac{s}{\sqrt{n}})],\ $will cover (overlay) the true mean $\theta
^{\ast }$, \textit{whatever that happens to be}, with probability $(1\mathbf{%
-}\alpha ),$ or equivalently, the error of coverage is $\alpha .$ Hence,
frequentist estimation the coverage error probability depends only on the
sampling distribution of $\overline{X}_{n}$ and is attached to random
interval for all values $\theta \mathbf{\neq }\theta ^{\ast }$ without
requiring one to know $\theta ^{\ast }.$

The evaluation at $\theta\mathbf{=}\theta^{\ast}$ calls into question the
decision-theoretic definition of unbiasedness:\vspace*{-0.06in}%
\[
\begin{tabular}{l}
$E(\widehat{\theta}_{n}(\mathbf{X}))\mathbf{=}\theta,$ $\forall\theta 
\mathit{\in}\Theta$,%
\end{tabular}
\vspace*{-0.06in}
\]
for frequentist estimation since this assertion makes no sense for all
values $\theta$ in $\Theta,$ but does make sense when defined at $\theta 
\mathbf{=}\theta^{\ast}.$ Similarly, the appropriate frequentist definition
of the MSE for an estimator, initially proposed by Fisher (1920), is defined
at the point $\theta\mathbf{=}\theta^{\ast}$:\vspace*{-0.06in}%
\begin{equation}
\begin{array}{c}
MSE(\widehat{\theta}_{n}(\mathbf{X});\theta^{\ast})\mathbf{=}E(\widehat{%
\theta }_{n}(\mathbf{X})\mathit{-}\theta^{\ast})^{2},\text{ for }%
\theta^{\ast }\mathit{\ }\text{in }\Theta.%
\end{array}
\vspace*{-0.06in}  \label{mse}
\end{equation}
Indeed, the well-known decomposition:$\vspace*{-0.06in}$%
\begin{equation}
\begin{array}{c}
MSE(\widehat{\theta}(\mathbf{X});\theta^{\ast})\mathbf{=}Var(\widehat{\theta 
}(\mathbf{X}))\mathit{+}[E(\widehat{\theta}_{n}(\mathbf{X}))\mathbf{-}%
\theta^{\ast}]^{2},\text{ for }\theta^{\ast}\mathit{\ }\text{in }\Theta,%
\end{array}
\vspace*{-0.06in}  \label{msf}
\end{equation}
is meaningful only when defined at the point $\theta\mathbf{=}\theta^{\ast}$%
(true mean) since by definition:$\vspace*{-0.06in}$ 
\begin{equation}
\begin{array}{l}
Var(\widehat{\theta}(\mathbf{X}))\mathit{=}E[\widehat{\theta}_{n}(\mathbf{X})%
\mathit{-}\theta_{m}]^{2},\ \theta_{m}\mathit{=}E(\widehat{\theta }_{n}(%
\mathbf{X}))\medskip \\ 
Bias(\widehat{\theta}_{n}(\mathbf{X});\theta^{\ast})\mathbf{=}E(\widehat{%
\theta}_{n}(\mathbf{X}))\mathbf{-}\theta^{\ast},\text{ }%
\end{array}
\vspace*{-0.06in}  \label{bias}
\end{equation}
and thus, the variance and the bias involve only two values of $\theta$ in $%
\Theta,$ $\theta_{m}$ and $\theta^{\ast},$ and when $\theta_{m}\mathit{=}%
\theta^{\ast}$ the estimator is unbiased. This implies that the apparent
affinity between the MSE$_{1}$ defined in (\ref{MSE}) and the variance of an
estimator is more apparent than real because the latter makes frequentist
sense only when $\theta_{m}\mathit{=}E(\widehat{\theta}_{n}(\mathbf{X}))$ is
a single point.\vspace*{-0.15in}

\subsection{James-Stein estimator from a frequentist perspective\protect%
\vspace*{-0.06in}}

For a proper frequentist evaluation of the above James-Stein result, it is
important to bring out the conflict between the \textit{overall MSE} (\ref%
{lm}) and the factual reasoning underlying frequentist estimation. From the
latter perspective, the James-Stein estimator raises several issues of
concern.

\textit{First,} both the Least-Squares $\widehat{\mathbf{\theta}}_{LS}(%
\mathbf{X})$ and the James-Stein $\widehat{\mathbf{\theta}}_{JS}(\mathbf{X}) 
$\ estimators are \textit{inconsistent} estimators of $\mathbf{\theta}$
since the underlying model suffers from the incidental parameter problem:
there is essentially one observation ($X_{k}$) for each unknown parameter ($%
\theta_{k}$), and as $m\rightarrow\infty$ the number of unknown parameters
increases at the same rate. To bring out the futility of comparing these two
estimators more clearly, consider the following simpler example.

\textbf{Example}. Let $\mathbf{X}\mathit{:=}(X_{1},X_{2},...,X_{n})$ be a
sample from the simple Normal model in (\ref{snm}). Comparing the two
estimators $\widehat{\theta }_{1}\mathbf{=}X_{n},\ \widehat{\theta }_{2}%
\mathbf{=}\frac{1}{2}\left( X_{1}\mathit{+}X_{n}\right) $ and inferring
that\ $\widehat{\theta }_{2}$ is relatively more efficient than $\widehat{%
\theta }_{1}$ relative to a square loss function, i.e.$\vspace*{-0.06in}$%
\[
\begin{array}{c}
\text{\textsf{MSE}}(\widehat{\theta }_{2}(\mathbf{X});\theta )\mathbf{=}1<%
\text{\textsf{MSE}}(\widehat{\theta }_{1}(\mathbf{X});\theta )\mathbf{=}%
\frac{1}{2},\ \forall \theta \mathit{\in }\mathbb{R},%
\end{array}%
\vspace*{-0.06in}
\]%
is totally uninteresting because both estimators are inconsistent!

\textit{Second}, to be able to discuss the role of admissibility in the
Stein (1956) result, we need to consider a consistent James-Stein estimator,
by extending the original data to a panel (longitudinal) data where the
sample is:\newline
$%
\begin{array}{c}
\mathbf{X}_{t}\mathbf{:=}(X_{1t},X_{2t},...,X_{mt}),\ t\mathbf{=}1,2,...,n.%
\end{array}
$ In this case the consistent Least-Squares and James-Stein estimators are:%
\vspace*{-0.06in}%
\[
\begin{array}{l}
\widehat{\mathbf{\theta}}_{LS}(\mathbf{X})\mathbf{=}\left( \overline{X}_{1},%
\overline{X}_{2},...,\overline{X}_{m}\right) ,\text{ where }\overline {X}_{k}%
\mathbf{=}\frac{1}{n}\sum\limits_{t=1}^{n}X_{kt},\ k\mathbf{=}1,2,...,m, \\ 
\widehat{\mathbf{\theta}}_{JS}^{+}(\mathbf{X})\mathbf{=}\left( 1-\frac {%
(m-2)\sigma^{2}}{\parallel\overline{\mathbf{X}}\parallel^{2}}\right) ^{+}%
\overline{\mathbf{X}},\ \text{where }\overline{\mathbf{X}}\mathbf{:=}\left( 
\overline{X}_{1},\overline{X}_{2},...,\overline{X}_{m}\right) .%
\end{array}
\vspace*{-0.06in}
\]
This enables us to evaluate the notion of `relatively better' more
objectively.

Admissibility relative to the overall loss function in (\ref{lm}) introduces
a trade-off between the accuracy of the estimators for individual parameters 
$\mathbf{\theta}\mathit{:=}(\theta_{1},\theta_{2},...,\theta_{m})$ and the
`overall' expected loss. The question is: `In what sense the overall MSE
among a group of mean estimates provides a better measure of `error' in
learning about the true values $\mathbf{\theta}^{\ast}\mathit{:=}%
(\theta_{1}^{\ast },\theta_{2}^{\ast},...,\theta_{m}^{\ast})$?' The short
answer is: it doesn't. Indeed, the overall MSE will be irrelevant when the
primary objective of estimation is to learn from data about $\mathbf{\theta}%
^{\ast}$. This is because the particular loss function penalizes the
estimator's capacity to pin-point $\mathbf{\theta}^{\ast}$ by trading an
increase in \textit{bias} for a decrease in the overall MSE in (\ref{lm}),
when the latter is misleadingly evaluated over all $\mathbf{\theta}$ in $%
\Theta\mathit{:=}\mathbb{R}^{m}$. That is, the James-Stein estimator flouts
the primary objective of pin-pointing $\mathbf{\theta}^{\ast}$ in favor of
reducing the overall MSE $\forall\mathbf{\theta}\mathit{\in}\Theta$.

In summary, the above discussion suggests that there is nothing paradoxical
about Stein's (1956) original result. What is problematic is not the
least-squares estimator, but the choice of `better' in terms of
admissibility relative to an overall MSE in evaluating the accuracy of the
estimators of $\mathbf{\theta}$.\vspace*{-0.15in}

\subsection{Frequentist hypothesis testing\protect\vspace*{-0.06in}}

Another frequentist inference procedure one can employ to learn from data
about $\theta^{\ast}$ is hypothesis testing, where the question posed is
whether $\theta^{\ast}$ is close enough to some prespecified value $\theta
_{0}$. In contrast to estimation, the reasoning underlying frequentist
testing is \textit{hypothetical }in nature. \vspace*{-0.15in}

\subsubsection{Legitimate frequentist error probabilities\protect\vspace*{%
-0.06in}}

For testing the hypotheses:\vspace*{-0.06in}%
\[
\begin{array}{c}
H_{0}\text{: }\theta\leq\theta_{0}\text{ vs. }H_{1}\text{: }\theta>\theta
_{0},\text{ where }\theta_{0}\text{ is a prespecified value,}%
\end{array}
\vspace*{-0.06in}
\]
one utilizes the same sampling distribution $\overline{X}_{n}\backsim $%
\textsf{N}$(\theta,\frac{1}{n})$, but transforms the pivot $d(\mathbf{X;}%
\theta)\mathit{:=}\sqrt{n}\left( \overline{X}_{n}\mathit{-}\theta^{\ast
}\right) $ into the test statistic by replacing $\theta^{\ast}$ with the
prespecified value $\theta_{0},$ yielding $d(\mathbf{X})\mathit{:=}\sqrt {n}%
\left( \overline{X}_{n}\mathit{-}\theta_{0}\right) .$ However, instead of
evaluating it under the factual $\theta\mathit{=}\theta^{\ast}$, it is now
evaluated under various \textit{hypothetical scenarios} associated with $%
H_{0}$ and $H_{1}$ to yield two types of (hypothetical) sampling
distributions:$\vspace*{-0.06in}$%
\[
\begin{tabular}{ll}
(I) & $d(\mathbf{X})\mathit{:=}\sqrt{n}\left( \overline{X}_{n}-\theta
_{0}\right) \overset{\theta=\theta_{0}}{\backsim}\text{\textsf{N}}%
(0,1),\medskip$ \\ 
(II) & $d(\mathbf{X})\mathit{:=}\sqrt{n}\left( \overline{X}_{n}\mathit{-}%
\theta_{0}\right) \overset{\theta=\theta_{1}}{\backsim }\text{\textsf{N}}%
(\delta_{1},1),\ \delta_{1}\mathit{=}\sqrt{n}\left( \theta_{1}\mathbf{-}%
\theta_{0}\right) $ for $\theta_{1}>\theta_{0}.$%
\end{tabular}
\vspace*{-0.06in}
\]
In both cases (I)-(II) the underlying reasoning is hypothetical in the sense
that the factual in (\ref{pq}) is replaced by hypothesized values of $%
\theta, $ and the test statistic $d(\mathbf{X})$ provides a standardized
distance between the hypothesized values ($\theta_{0}$ or $\theta_{1}$) and $%
\theta^{\ast}$ the true $\theta,$ assumed to underlie the generation of the
data $\mathbf{x}_{0},$ yielding $d(\mathbf{x}_{0}).$ Using the sampling
distribution in (I) one can define the following legitimate error
probabilities:$\vspace*{-0.06in}$%
\begin{equation}
\begin{tabular}{rl}
\textbf{significance level:} & $\mathbb{P}(d(\mathbf{X})>c_{\alpha};H_{0})=%
\alpha,\medskip$ \\ 
\textbf{p-value:} & $\mathbb{P}(d(\mathbf{X})>d(\mathbf{x}_{0});H_{0})%
\mathbf{=}p(\mathbf{x}_{0}).$%
\end{tabular}
\vspace*{-0.06in}  \label{I}
\end{equation}
Using the sampling distribution in (II) one can define:$\vspace*{-0.06in}$%
\begin{equation}
\hspace*{-0.5in}%
\begin{tabular}{rl}
\textbf{type II error prob.:} & $\mathbb{P}(d(\mathbf{X})\mathit{\leq }%
c_{\alpha};\theta\mathbf{=}\theta_{1})\mathbf{=}\beta(\theta_{1}),$\ for $%
\theta_{1}\mathit{>}\theta_{0},\medskip$ \\ 
\textbf{power:} & $\mathbb{P}(d(\mathbf{X})\mathit{>}c_{\alpha};\theta 
\mathbf{=}\theta_{1})\mathbf{=}\varrho(\theta_{1}),$\ for $\theta _{1}%
\mathit{>}\theta_{0}.$%
\end{tabular}
\ \vspace*{-0.06in}  \label{II}
\end{equation}

It can be shown that the test $T_{\alpha},$ defined by the test statistic $d(%
\mathbf{X})$ and the rejection region $C_{1}(\alpha)\mathbf{=}\{\mathbf{x:}d(%
\mathbf{x})>c_{\alpha}\},$ constitutes a Uniformly Most Powerful (UMP) test
for significance level $\alpha;$ see Lehmann (1959). The type I [II] error
probability is associated with test $T_{\alpha}$ erroneously rejecting
[accepting] $H_{0}$. The type I and II error probabilities evaluate the
generic capacity [whatever the sample realization $\mathbf{x}\mathit{\in }%
\mathbb{R}^{n}$] of a test to reach correct inferences. Contrary to Bayesian
claims, these error probabilities have nothing to do with the temporal or
the physical dimension of the long-run metaphor associated with repeated
samples. The relevant feature of the long-run metaphor is the repeatability
(in principle) of the DGM represented by $\mathcal{M}_{\mathbf{\theta}}(%
\mathbf{x});$ this feature can be easily operationalized using computer
simulation; see Spanos (2013).

The key difference between the significance level $\alpha$ and the p-value
is that the former is a \textit{pre-data} and the latter a \textit{post-data}
error probability. Indeed, the p-value can be viewed as the smallest
significance level $\alpha$ at which $H_{0}$ would have been rejected with
data $\mathbf{x}_{0}$. The legitimacy of post-data error probabilities
underlying the hypothetical reasoning can be used to go beyond the N-P
accept/reject rules and provide an evidential interpretation pertaining to
the discrepancy $\gamma$ from the null warranted by data $\mathbf{x}_{0}$;
see Mayo and Spanos (2006).

Despite the fact that frequentist testing uses hypothetical reasoning, its
main objective is also to learn from data about the true model $\mathcal{M}%
^{\ast}(\mathbf{x})\mathbf{=}\{f(\mathbf{x};\theta^{\ast})\},\ \mathbf{x}%
\mathit{\in}\mathbb{R}_{X}^{n}.$ This is because a test statistic like $d(%
\mathbf{X})\mathbf{:=}\sqrt{n}\left( \overline{X}_{n}\mathbf{-}\theta
_{0}\right) $ constitutes nothing more than a scaled distance between $%
\theta^{\ast}\ [$the value behind the generation of $\overline{x}_{n}],$ and
a hypothesized value $\theta_{0},$ with $\theta^{\ast}$ being replaced by
its `best' estimator $\overline{X}_{n}.$ \vspace*{-0.15in}

\section{Revisiting loss and risk functions\protect\vspace*{-0.06in}}

The above discussion raises serious questions about the role of loss
functions and admissibility in evaluating learning from data $\mathbf{x}_{0}$
about $\mathbf{\theta}^{\ast}.$ In particular:

(i) What does the extraneous information concerning costs associated with
different parameter values have to do with learning about $\mathbf{\theta }%
^{\ast}$?

(ii) In what sense is an inconsistent but relatively (to a particular loss
function) efficient an `optimal' estimator for learning about $\mathbf{%
\theta }^{\ast}$?

(iii) Why is the overall MSE more important than learning from data about
the true values of $\mathbf{\theta}^{\ast}?$ \vspace*{-0.12in}

\subsection{Where do loss functions come from?\protect\vspace*{-0.06in}}

A closer scrutiny of the decision-theoretic set up reveals that the loss
function needs to invoke `information from sources other than the data',
which is usually not readily available. Indeed, such information is
available in very restrictive situations, such as acceptance sampling in
quality control. In light of that, a proper understanding of the intended
scope of statistical inference calls for distinguishing the special cases
where the loss function is part and parcel of the available substantive
information from those that no such information is either relevant or
available.

Tiao and Box (1975), p. 624, reiterated Fisher's (1935) distinction:

\textsf{\textquotedblleft Now it is undoubtedly true that on the one hand
that situations exist where the loss function is at least approximately
known (for example certain problems in business) and sampling inspection are
of this sort. ... On the other hand, a vast number of inferential problems
occur, particularly in the analysis of scientific data, where there is no
way of knowing in advance to what use the results of research will
subsequently be put.\textquotedblright}

Cox (1978), p. 45, went further and questioned this framing even in cases
where the inference might involve a decision:

\hspace*{-0.25in}\textsf{\textquotedblleft The reasons that the detailed
techniques [decision-theoretic] seem of fairly limited applicability, even
when a fairly clear cut decision element is involved, may be (i) that,
except in such fields as control theory and acceptance sampling, a major
contribution of statistical technique is in presenting the evidence in
incisive form for discussion, rather than in providing mechanical
presentation for the final decision. This is especially the case when a
single major decision is involved. (ii) The central difficulty may be in
formulating the elements required for the quantitative analysis, rather than
in combining these elements via a decision rule.\textquotedblright}

Another important aspect of using loss functions in inference is that in
practice they seem to be an add-on to the inference itself since they bring
to the problem the information other than the data. In particular, the same
statistical inference problem can give rise to very different
decisions/actions depending on one's loss function. To illustrate that
consider an example from Chatterjee (2002):

\textsf{\textquotedblleft... consider the case of a new drug whose effects
are studied by a research scientist attached to the laboratory of a
pharmaceutical company. The conclusion of the study may have different
bearings on the action to be taken by (a) the scientist whose line of
further investigation would depend on it, (b) the company whose business
decisions would determined by it, and (c) the Government whose policies as
to health care, drug control, etc. would take shape on that
basis.\textquotedblright\ (p. 72) }

In practice, each one of these different agents is likely to have a very
different loss function, but their inferences have a common denominator: the
scientific evidence which relating to the true $\theta$ that stems solely
from the observed data?

Finally, the extreme relativism of loss function optimality renders
decision-theoretic and Bayes rules highly vulnerable to abuse. In practice,
one can justify any estimator as optimal, however lame in terms of other
criteria, by selecting the "appropriate" loss function.

\textbf{Example}. Consider a manufacturer of high precision bolts and nuts
who has information that the buyer only checks the first and last box for
quality control when accepting an order. This suggests that to minimize
losses, stemming from the return of its products as defective, an
appropriate loss function might be:$\vspace*{-0.06in}$%
\begin{equation}
\begin{array}{c}
L(\mathbf{X};\theta )\mathit{=}\left( [(X_{1}\mathit{+}X_{n})/2]-\theta
\right) ^{2},\ \theta \mathit{\in }(0,1).%
\end{array}%
\vspace*{-0.06in}  \label{2l}
\end{equation}%
The `optimal' estimator relative to (\ref{2l}) is $\widetilde{\theta }%
\mathit{=}(X_{1}\mathit{+}X_{n})/2$, but $\widetilde{\theta }$ is a terrible
estimator for pinpointing $\theta ^{\ast }\ $because it is inconsistent!%
\vspace*{-0.15in}

\subsection{Loss functions vs. inherent distance functions\protect\vspace*{%
-0.07in}}

The notion of a loss function stemming from `information other than the
data' raises another source of potential conflict. This stems from the fact
that within each statistical model $\mathcal{M}_{\mathbf{\theta}}(\mathbf{x}%
) $ there exists an \textit{inherent} statistical distance function, often
relating to the log-likelihood and the score function, and hence stemming
from information contained in the data; see Casella and Berger (2002).

It is well-known that when the distribution underlying $\mathcal{M}_{\mathbf{%
\theta}}(\mathbf{x})$ is Normal, the \textit{inherent distance function} for
comparing estimators of the mean ($\theta$) is the square:$\vspace*{-0.08in}$%
\[
\begin{array}{c}
ND(\widehat{\theta}_{n}(\mathbf{X});\theta^{\ast})\mathit{=}(\widehat{\theta 
}_{n}(\mathbf{X})-\theta^{\ast})^{2}.%
\end{array}
\vspace*{-0.08in}
\]
On the other hand, when the distribution is \textit{Laplace\ }the relevant
statistical distance function is the \textit{Absolute Distance }(see Shao,
2003):$\vspace*{-0.08in}$%
\[
\begin{array}{c}
AD(\widehat{\theta}_{n}(\mathbf{X});\theta^{\ast})=|\widehat{\theta}_{n}(%
\mathbf{X})-\theta^{\ast}|.%
\end{array}
\vspace*{-0.08in}
\]
Similarly, when the distribution underlying $\mathcal{M}_{\mathbf{\theta}}(%
\mathbf{x})$ is \textit{Uniform}, the inherent distance function is:$%
\vspace*{-0.08in}$%
\[
\begin{array}{c}
SUP(\widehat{\theta}_{n}(\mathbf{X});\theta^{\ast})=\underset{\mathbf{x}\in%
\mathbb{R}_{X}^{n}}{\sup}|\widehat{\theta}_{n}(\mathbf{x})-\theta^{\ast}|.%
\end{array}
\vspace*{-0.1in}
\]
A key feature of all these distance functions is that they are defined at
the point $\theta\mathbf{=}\theta^{\ast}$ and not for all $\theta$ in $%
\Theta $, as the traditional loss functions.

The question that naturally arises is when it might make sense to ignore
these inherent distance functions and compare estimators using an externally
given loss function. The key difference between the two is that any
assumptions that comprise the likelihood function are testable vis-a-vis the
data, but those underlying the loss function are not. Moreover, the
likelihood function gives rise to a `global' notion of optimality, known as 
\textit{full efficiency} defined at $\mathbf{\theta }\mathit{=}\mathbf{%
\theta }^{\ast }$ in terms of Fisher's information:$\vspace*{-0.08in}$%
\[
\begin{array}{c}
CR(\mathbf{\theta }^{\ast })\mathbf{=}\mathbb{I}_{n}^{-1}(\mathbf{\theta }%
^{\ast }),\text{ }\mathbb{I}_{n}(\mathbf{\theta }^{\ast })\mathit{:=}E\left(
-\frac{\partial ^{2}\ln L(\mathbf{\theta })}{\partial \mathbf{\theta }%
\partial \mathbf{\theta }^{\mathbf{\top }}}\right) .%
\end{array}%
\vspace*{-0.08in}
\]%
What is an optimal estimator depends only on the information contained in
the statistical model $\mathcal{M}_{\mathbf{\theta }}(\mathbf{x})$. This
contrasts with admissibility which is a property defined in terms of `local'
optimality \textit{relative} to a loss function based on outside information
and evaluated $\forall \mathbf{\theta }\mathit{\in }\mathbb{R}^{m}$. 
\vspace*{-0.15in}

\subsection{Decisions vs. inferences\protect\vspace*{-0.06in}}

The above discussion brings out the crucial distinction between a `decision'
and an `inference' stemming from data $\mathbf{x}_{0}$.

Even before Wald (1939) introduced the decision-theoretic perspective,
Fisher (1935) perceptively argued:

\textsf{\textquotedblleft In the field of pure research no assessment of the
cost of wrong conclusions, or of delay in arriving at more correct
conclusions can conceivably be more than a pretence, and in any case such an
assessment would be inadmissible and irrelevant in judging the state of the
scientific evidence.\textquotedblright\ (pp. 25-26)}

Tukey (1960) echoed Fisher's view by contrasting decisions vs. inferences:

\textsf{\textquotedblleft Like any other human endeavor, science involves
many decisions, but it progresses by the building up of a fairly well
established body of knowledge. This body grows by the reaching of
conclusions -- by acts whose essential characteristics differ widely from
the making of decisions. Conclusions are established with careful regard to
evidence, but without regard to consequences of specific actions in specific
circumstances.\textquotedblright\ (p. 425)}

Tukey also recognized how decision theory distorts frequentist testing by
replacing error probabilities with losses and costs:

\textsf{\textquotedblleft Wald's decision theory ... has given up fixed
probability of errors of the first kind, and has focused on gains, losses or
regrets.\textquotedblright\ (p. 433)}

Hacking (1965) brought out the key difference between an `inference
pertaining to evidence' for or against a hypothesis, and a `decision to do
something' as a result of an inference:

\textsf{\textquotedblleft... to conclude that an hypothesis is best
supported is, apparently, to decide that the hypothesis in question is best
supported. Hence it is a decision like any other. But this inference is
fallacious. Deciding that something is the case differs from deciding to do
something. ... Hence deciding to do something falls squarely in the province
of decision theory, but deciding that something is the case does
not.\textquotedblright\ (p. 31)}

This issue was elaborated upon by Birnbaum (1977), p. 19:

\textsf{\textquotedblleft Two contrasting interpretations of the decision
concept are formulated: \textit{behavioral}, applicable to `decisions' in a
concrete literal sense as in acceptance sampling; and\hspace*{-0.25in}%
\textit{evidential}, applicable to `decisions' such as 'reject }$H_{0}$%
\textsf{' in a research context, where the pattern and strength of
statistical evidence concerning statistical hypotheses is of central
interest.\textquotedblright}\vspace*{-0.15in}

\subsection{Acceptance sampling vs. learning from data\protect\vspace*{%
-0.06in}}

Let us bring out the key features of a situation where the above
decision-theoretic set up makes perfectly good sense. This is the situation
Fisher (1955) called \textit{acceptance sampling}, such as an industrial
production process where the objective is quality control, i.e. to make a
decision pertaining to shipping sub-standard products (e.g. nuts and bolts)
to a buyer using the expected loss/gain as the ultimate criterion.

In an acceptance sampling context, the \textsf{MSE}$(\widehat{\theta }(%
\mathbf{X});\theta),$ or some other risk function, are relevant because they
evaluate genuine losses associated with a decision related to the choice of
an estimate $\widehat{\theta}(\mathbf{x}_{0})$, say the cost of the observed
percentage of defective products, but that has nothing to do with type I and
II error probabilities.

Acceptance sampling differs from a scientific enquiry in two crucial
respects:

\textbf{[a]} The primary aim is to use statistical rules to guide actions
astutely, e.g. use $\widehat{\theta}(\mathbf{x}_{0})$ in order to minimize
the expected loss associated with \textquotedblleft a
decision\textquotedblright, and

\textbf{[b]} The sagacity of all actions is determined by the respective
`losses' stemming from \textquotedblleft\textit{relevant information\ other
than the data}\textquotedblright\ (Cox and Hinkley, 1974, p. 251).

The key difference between acceptance sampling and a scientific inquiry is
that the primary objective of the latter is \textit{not} to minimize
expected loss (costs, utility) associated with different values of $\mathbf{%
\theta }\mathit{\in}\Theta,$ but to use data $\mathbf{x}_{0}$ to learn about
the `true' model (\ref{dgm}). The two situations are drastically different
mainly because the key notion of a `true $\mathbf{\theta}$' calls into
question the above acceptance sampling set up. Indeed, the loss function
being defined `$\forall\mathbf{\theta}\mathit{\in}\Theta$', will penalize $%
\mathbf{\theta }^{\ast},$ since there is no reason to believe that the
lowest ranked $\mathbf{\theta}$ would coincide with $\mathbf{\theta}^{\ast}$%
, unless by accident.

Consider the case where acceptance sampling resembles hypothesis testing in
so far as final products are randomly selected for inspection during the
production process. In such a situation the main objective can be viewed as
operationalizing the probabilities of false acceptance/rejection with a view
to minimize the expected losses. The conventional wisdom has been that this
situation is similar enough to Neyman-Pearson (N-P) testing to render the
latter as the appropriate framing for the decision to ship this particular
batch or not. However, a closer look at some of the examples used to
illustrate such a situation (Silvey, 1975), reveals that the decisions are
driven exclusively by the risk function and not by any quest to learn from
data about the true $\mathbf{\theta}^{\ast}$. For instance, N-P way of
addressing the trade-off between the two types of error probabilities,
fixing $\alpha$ to a small value and seek a test that minimizes the type II
error probability, seems utterly irrelevant in such a context. One can
easily think of a loss function where the `optimal' trade-off calls for a
much larger type I than type II error probability.

In light of the above discussion, what is different in acceptance sampling
is that:

\textbf{[c]} The trade-off between the two types of error probabilities is
determined by the risk function itself, and not by any attempt to learn from
data about $\mathbf{\theta}^{\ast}.$ Indeed, this learning is deliberately
undermined by certain loss function such as the \textit{overall} MSE (\ref%
{lm}) that favor biased estimators of the James-Stein type.

Given the crucial differences in \textbf{[a]-[c]}, one can make a strong
case that the objectives and the underlying reasoning of acceptance sampling
are drastically different from those pertaining to learning from data in a
scientific context.\vspace*{-0.15in}

\subsection{Is expected loss a legitimate frequentist error?\protect\vspace*{%
-0.06in}}

The key question is whether expected loss is a legitimate frequentist error
like bias, MSE and the type I-II error. `What do these legitimate
frequentist errors have in common?'

\textit{First}, they stem directly from the statistical model $\mathcal{M}_{%
\mathbf{\theta}}(\mathbf{x})$ since the underlying sampling distributions of
estimators, test statistics and predictors are derived \textit{exclusively
from the distribution of the sample} $f(\mathbf{x};\mathbf{\theta})$ via (%
\ref{f}). In this sense, the relevant error probabilities are directly
related to statistical information pertaining to the data as summarized by
the statistical model $\mathcal{M}_{\mathbf{\theta}}(\mathbf{x})$ itself.

\textit{Second}, they are attached to a particular frequentist inference
procedure as they relate to a relevant inferential claim. These error
probabilities calibrate the effectiveness of inference procedures in
learning from data about the true statistical model $\mathcal{M}^{\ast}(%
\mathbf{x})\mathbf{=}\{f(\mathbf{x};\mathbf{\theta}^{\ast})\},\ \mathbf{x}%
\mathit{\in }\mathbb{R}_{X}^{n}.$

In light of these features, the question is: `in what sense a risk function
could potentially represent relevant frequentist errors?' That argument that
the risk function represents legitimate frequentist errors because it is
derived by taking expectations with respect to $f(\mathbf{x};\theta),$ $%
\mathbf{x}\mathit{\in}\mathbb{R}_{X}^{n}$ (Robert (2001), is misguided for
two reasons.

(a) The relevant errors in estimation, including the bias $E(\widehat{\theta 
}_{n}(\mathbf{X}))\mathbf{-}\theta^{\ast}$ and MSE $E(\widehat{\theta}_{n}(%
\mathbf{X})\mathbf{-}\theta^{\ast})^{2},$ are evaluated with respect to $f(%
\mathbf{x};\theta^{\ast}),$ $\mathbf{x}\mathit{\in}\mathbb{R}_{X}^{n}$, by
invoking factual reasoning ($\theta^{\ast}$ is assumed to be the state of
Nature). Wald's (1939) original loss function in (\ref{w}) represents an
interesting case because it is defined in terms of $\theta^{\ast}$, which
renders it non-operational when evaluated for all $\theta$ in $\Theta$,
since $\theta^{\ast}$ is unknown in practice. In contrast, the errors
associated with the bias and MSE are rendered operational by the factual
reasoning fashioned to forgo knowing $\theta^{\ast}$.

(b) The expected losses stemming from the risk function $R(\theta ,\widehat{%
\theta})$ are attached to particular values of $\theta$ in $\Theta$. Such an
assignment is in direct conflict with all the above legitimate error
probabilities that are attached to the inference procedure itself, and never
to the particular values of $\theta$ in $\Theta.$ The expected loss assigned
to each value of $\theta\mathbf{\ }$in $\Theta$ has nothing to do with
learning from data about $\theta^{\ast}$. Indeed, the risk function will
penalize a procedure for pin-pointing\textbf{\ }$\theta^{\ast}$ since the
latter is unknown in practice. This is in direct conflict with the main
objective of frequentist estimation but in sync with `acceptance sampling',
where the objective of the inference has everything to do with expected
losses.\vspace*{-0.15in}

\section{Summary and conclusions\protect\vspace*{-0.06in}}

The paper makes a case for Fisher's (1935; 1955) assertions concerning the
appropriateness of the decision-theoretic framing for `acceptance sampling'
and its inappropriateness for frequentist inference. A closer look at this
framing reveals that it is congruent with the Bayesian approach because
provides it with a theory of optimal inference. Decision-theoretic and
Bayesian rules are considered optimal when they minimize the expected loss
for all possible values of $\theta $ [$\forall \theta \mathit{\in }\Theta ],$
irrespective of what the true value $\theta ^{\ast }$ happens to be. In
contrast, the theory of optimal frequentist inference revolves around the
true value $\theta ^{\ast }$, since it depends entirely on the capacity of
the procedure to pinpoint $\theta ^{\ast }.$ The frequentist approach relies
on factual (estimation, prediction), as well as hypothetical (testing)
reasoning, both of which revolve around the existential quantifier $\exists
\theta ^{\ast }\mathit{\in }\Theta $. The inappropriateness of the
quantifier $\forall \theta \mathit{\in }\Theta $ calls into question the
relevance of \textit{admissibility} as a minimal property for frequentist
estimators. A strong case can be made that the relevant minimal property for
frequentist estimators is \textit{consistency}. In addition, full efficiency
provides the relevant measure of an estimator's finite sample efficiency
(accuracy) in pinpointing $\theta ^{\ast }$. Both of these properties stem
from the underlying statistical model $\mathcal{M}_{\mathbf{\theta }}(%
\mathbf{x}),$ in contrast to admissibility which relies on loss functions
based on information other than the data.

It is argued that Stein's (1956) result stems from the fact that
admissibility introduces a trade-off between the accuracy of the estimator
in pinpointing $\mathbf{\theta }^{\ast }$ and the `overall' expected loss.
That is,\ the James-Stein estimator achieves a higher overall MSE by
blunting the capacity of a frequentist estimator to pinpoint $\mathbf{\theta 
}^{\ast }.$ Why would a frequentist care about the overall MSE defined for
all $\mathit{\theta }$ in $\Theta ?$ After all, expected losses are not
legitimate errors similar to bias and MSE (when properly defined), as well
as coverage, type I and II errors. The latter are attached to the
frequentist procedures themselves to calibrate their capacity to achieve
learning from data about $\mathbf{\theta }^{\ast }$. In contrast, expected
losses are assigned to different values of $\mathbf{\theta }$ in $\Theta $,
using information other than the data. 

\end{document}